\documentclass[12pt, fleqn]{article}
\usepackage[cp1251]{inputenc}
\usepackage{latexsym,amsfonts,amssymb}
\usepackage{graphicx}

\usepackage{amsbsy}
\usepackage{amsmath}
\usepackage{epsf}
\usepackage{cite}

\newtheorem{theorem}{Theorem}

\newtheorem{remark}{Remark}
\newcommand{\bt}{\begin{theo}}
\newcommand{\et}{\end{theo}}
\newcommand{\bd}{\begin{displaymath}}
\newcommand{\ed}{\end{displaymath}}

\newcommand{\rg}{\right}

\newcommand{\be} {\begin{equation}}
\newcommand{\ee} {\end{equation}}
\newcommand{\ba} {\begin{array}}
\newcommand{\ea} {\end{array}}
\newcommand{\bea}{\begin{eqnarray}}
\newcommand{\eea} {\end{eqnarray}}

\newcommand{\p} {\partial}

\newcommand{\lbd} {\lambda}

\begin{document}

\begin{center}
 {\Large \bf
An age-structured diffusive  model for epidemic \\ modelling: Lie
symmetries and exact solutions  }

\medskip

{\bf Roman Cherniha $^{a,b,}$\footnote{\small  Corresponding author.
E-mail: r.m.cherniha@gmail.com; roman.cherniha1@nottingham.ac.uk}
and   Vasyl' Davydovych$^{c}$ }

$^{a}$  \quad National University of Kyiv-Mohyla Academy, 2,
Skovoroda Street, Kyiv  04070, Ukraine

$^{b}$ \quad School of Mathematical Sciences, University of Nottingham,\\
  University Park, Nottingham NG7 2RD, UK

$^{c}$ \quad Institute of Mathematics,  National Academy of Sciences
of Ukraine, \\
 3, Tereshchenkivs'ka Street, Kyiv 01004, Ukraine\\

\end{center}

\begin{abstract}
A new age-structured diffusive  model
 for the   mathematical modelling of epidemics is suggested. The model  can be considered as a generalization of  two  models
 suggested earlier for similar purposes. The Lie symmetry classification of the model is  derived. It is shown that the model admits an infinite-dimensional Lie algebra of invariance.    Using the Lie symmetries,
 exact solutions, in particular those of the travelling wave types and in terms of special functions, are   constructed. Examples of application of  exact solutions with the correctly-specified parameters  for calculation of the total number  of infected individuals during an epidemic  are presented.

\end{abstract}

\section{Introduction } \label{sec-1}

A mathematical model describing the spread of the COVID-19 pandemic
was suggested   in \cite{ch-dav-2020} and studied in detail in
\cite{ch-dav-2022-cd,ch-du-da-2024}.
 The model is a generalization of a simpler   model based on ordinary differential equations (see \cite{ch-dav-2020})
  that was successfully  used for modelling the first wave of the COVID-19  in some countries.
   The generalization takes into account spacial heterogeneity  by introducing     diffusion terms and reads as
\begin{equation}\label{2-1}\begin{array}{l} \medskip  u_t = d_1 \Delta u+u(a- bu^\gamma),  \\
v_t = d_2\Delta u +k(t)u.\end{array}\end{equation}
In (\ref{2-1}), the function $u(t,x,y)$ describes the density (rate)
of the infected persons (the number of the  COVID-19 cases) in a
vicinity of the point $(x,y)$,  while $v(t,x,y)$ means the density
of the deaths from COVID-19. The diffusivity coefficients $d_1$ and
$d_2$ describe the random movement  of the infected persons, which
lead to increasing the pandemic spread. Each  coefficient in the
reactions terms, $a, \ b, \ \gamma$  and $k(t)$, has the clear
meaning described in \cite{ch-dav-2020, ch-dav-2022-cd}. Recently
\cite{ch-du-da-2024}, it was demonstrated that the mathematical
model with the governing equations (\ref{2-1}) and relevant boundary
and initial conditions adequately describes the second COVID-19
pandemic wave in Ukraine. A rigorous comparison of the numerical
results obtained by numerical simulations  with the official data
proved  that the model produces very plausible total numbers of the
Covid-19 cases and deaths. An extensive analysis of impact of the
parameters arising in the model is also  presented in
\cite{ch-du-da-2024}.

On the other hand, it is well-known that age of infective
individuals has an essential impact on recovering rate.
 Generally speaking, old individuals have less chances to recover comparing with young those, therefore the death
 rate is growing with the age.
The simplest epidemic  model taking into account the age reads as
\begin{equation}\label{2-2}
u_t+\mu_0u_{\tau}= -\mu(\tau)u,
\end{equation} where $\mu(\tau)$ is a death rate and $\mu_0>0$ is a
parameter (typically $\mu_0=1$ is taken but this assumption is
questionable). Following the pioneering study \cite{foerster-59},
equation (\ref{2-2}) is also called the Von Foerster equation. This
linear equation was extensively studied many years ago
\cite{trucco-65,trucco-1965,murray-1989}. We also note that there
are recent studies, in which generalizations of (\ref{2-2})
involving integral terms are suggested (see
\cite{bentout-21,McCluskey-16} and papers cited therein).

Later,  the epidemic model (\ref{2-2}) was  extended by taking into
account the  diffusion effect. As a result, the reaction-diffusion
equation  \cite{gopalsamy-77,gurtin-81,gurtin-82,metz-86}
\begin{equation}\label{2-7} u_t+\mu_0u_{\tau} = d_1 \Delta u - \mu(\tau) u
\end{equation}
was suggested as a natural generalization of the Von Foerster
equation.

 At the present time, the existence of  solutions,
including traveling waves and periodic solutions,  of  model
(\ref{2-7}) and its generalizations was  studied in many works (see,
e.g., \cite{kubo-91,kubo-07,jin-09,liu-21}).
Obviously, the reaction-diffusion equation (\ref{2-7}) is linear,
therefore many classical techniques can directly be applied for its
solving and rigorous analysis.

On the other hand, almost all known space distributed  models used
for modelling in epidemiology  are based on nonlinear
reaction-diffusion equations (see the recent review
\cite{da-da-ch-23} and citations therein). In our opinion, there is
some contradiction that the mathematical models used for the
epidemic spread, but neglecting  age of a population,  are more
complicated  than those  taking into account age structure of the
population in question.  Motivated by this observation, we suggest
here a nonlinear generalization of the linear PDE (\ref{2-7}), study
its Lie symmetry, construct exact solutions and suggest their
interpretation.

The reminder of this paper  is organized as follows. In
Section~\ref{sec-2},  a nonlinear generalization of equation
(\ref{2-7}) is presented and analyzed. In Section~\ref{sec-3}, the
Lie symmetry classification of the generalized equation is derived.
In Section~\ref{sec-4}, a wide range of exact solutions, including
travelling  waves and  those involving special functions, are
constructed and their properties are briefly discussed. An
illustrative example of application of the correctly-specified
exact solution  for calculation of  total numbers  of infected
individuals during an epidemic is presented as well. Finally, we
discuss the results obtained and present some conclusions  in the
last section.

\section{An age-structured model with diffusion} \label{sec-2}

First of all, let us consider equation (\ref{2-7}). One may easily
observe that the assumptions $u_{\tau}=0$ and $\mu(\tau)=\mu_0={\rm
const}$, i.e.  the epidemic spread does not depend on age of the
given  population, immediately leads to the equation
\begin{equation}\label{2-7*} u_t  = d_1 \Delta u - \mu_0 u.
\end{equation}
The latter is the classical diffusion equation with the sink $\mu_0
u$. To the best of our knowledge, equation (\ref{2-7*}) does not
occur for the mathematical modelling of epidemics. However, if  we
combine the first equation from (\ref{2-1}) with (\ref{2-2}) then
the following
 nonlinear PDE  is obtained:
\begin{equation}\label{2-3}
u_t +\mu_0u_{\tau}= d_1 \Delta u+(a-\mu(\tau))u- bu^{\gamma+1},
\end{equation}
where $u(t,\tau,x,y)$ describes the density (rate) of the infected
persons of age $\tau$ in a vicinity of the point $(x,y)$.
 We remind the reader that  $a>0$ is  the
coefficient for the virus transmission mechanism, $b>0$ is  the
coefficient for the effectiveness of government restrictions
(quarantine rules), $\gamma>0$ is the  exponent,  which guarantees
that  the total number of the infected persons   is bounded in time,
and its value depends mostly on the country (large region) in
question \cite{ch-dav-2020}.

Thus, the total number of infected  individuals of the age $\tau $
up to time $t$ is  calculated using the formula
\[ U(t,\tau)=
\underset{\Omega}\int u(t,\tau,x,y)dxdy, \] where $\Omega \subset
\textbf{R}^2 $ is a domain (continent/country/region) in which a
pandemic is spread. The total number of infected population is
obtained by the double integral
\begin{equation}\label{2-4*} U_{\rm total}(t)=\underset{I}\int\left(\underset{\Omega}\int u(t,\tau,x,y)dxdy\right)d\tau, \end{equation}
where $I=[0,\tau_{\max}]$,  $\tau_{\max}$ is a maximum life span in
population.

Equation  can be rewritten in the form
\[u_t +\mu_0u_{\tau}= d_1 \Delta u-\mu^*(\tau)u- bu^{\gamma+1} \]
by introducing the notation $\mu^*(\tau)=\mu(\tau)-a$. Moreover, as
it  was shown in \cite{ch-du-da-2024},  one should take  nonconstant
(in space) coefficients in  the governing equations of (\ref{2-1})
in order to ideally adopt the model to the official data of the
COVID-19 pandemic. Thus, a natural generalization of (\ref{2-3})
reads as
\begin{equation}\label{2-4}
u_t +\mu_0u_{\tau}= d_1 \Delta u+ A(\tau,x,y)u- B(x,y)u^{\gamma+1},
\end{equation}
 where  $A(\tau,x,y)$ and $B(x,y)$ are given  functions.

 It is well known that epidemics spread much faster in  agglomerations. It means that the coefficient $b$ reflecting
  the effectiveness of the government restrictions (quarantine rules) takes the smallest value in the
  center of the agglomeration, say, a city  with the coordinates $(x,y)=(0,0)$. Notably, we have  shown  this for
    the Kyiv city  agglomeration   in the case of the COVID-19 pandemic in Ukraine \cite{ch-du-da-2024}. Thus, one may expect
 that   $B(x,y)= b_0 +b_1(x,y) $, where $b_0$ is a nonnegative parameter and
  $b_1(x,y)$ is  a nonnegative function satisfying the condition $b_1(0,0)=0$
  that increases with  growing  $x^2+y^2$. A simplest example could be \[b_1(x,y)=b_1(x^2+y^2)^\kappa, \ \kappa>0, \
  b_1>0.\]
  As it is shown in  \cite{ch-du-da-2024}, the  coefficient $a$ reflecting the virus transmission mechanism
  does not depend so much on the  population density. So, one may take  $A(\tau,x,y)=a_0 + \varepsilon(x^2+y^2)^\nu- \mu(\tau)  $
  with $\nu>0, \ a_0 >0$ and $|\varepsilon|\ll 1$. As a result, a plausible example of the general model (\ref{2-4}) has the form
  \begin{equation}\label{2-5*} u_t +\mu_0u_{\tau}= d_1 \Delta u+ \Big(a_0 + \varepsilon(x^2+y^2)^\nu-
  \mu(\tau)\Big)u- \Big(b_0 + b_1(x^2+y^2)^\kappa\Big)u^{\gamma+1}.  \end{equation}
  Obviously, the latter reduces to  the form (\ref{2-3}) provided $\varepsilon=b_1=0$.
  Another interesting case for possible applications occurs when $\varepsilon=b_0=0$.

 In what follows, we consider  (\ref{2-3}) as a basic equation for age-structured models with diffusion.
The nonlinear reaction-diffusion-convection equation (\ref{2-3}) can
be simplified provided natural assumptions about positivity of
parameters $\mu_0, \ a, \ b$ and $d_1$ take place (here we do not
consider  limiting case when some of them  are zero). Let us apply
the scale transformations
\begin{equation}\label{2-5}
t^*= at,\ \tau^*=\frac{a}{\mu_0}\,\tau, \ \mu^*= \frac{1}{a}\,\mu, \
x^*=\sqrt{\frac{a}{d_1}}\,x, \ y^*=\sqrt{\frac{a}{d_1}}\,y, \ u^*=
\left(\frac{b}{a}\right)^{1/\gamma}u
\end{equation} in order to
simplify equation (\ref{2-3}). Thus, omitting stars $*$, one obtains
the following nonlinear PDE involving only the parameter $\gamma$
and the function $\mu(\tau)$:
\begin{equation}\label{2-6}
u_t +u_{\tau}=  \Delta u+(1-\mu(\tau))u- u^{\gamma+1}.
\end{equation}
The nonlinear PDE (\ref{2-6}) is the main object of this study.

\section{Lie symmetry classification of the model } \label{sec-3}

In this section, we study Lie symmetries of equation (\ref{2-6})
with an arbitrary number of space variables: \be\label{3-1}
u_t+u_{\tau}=\Delta u+(1-\mu(\tau))u-u^{\gamma+1}, \ee where
$u(t,\tau,x)$ is an unknown function of $n+2$ variables $t, \ \tau$,
\  $x=(x_1,\dots,x_n)$; \   $\Delta= \frac{\p^2}{\p x_1^2}+\dots
+\frac{\p^2}{\p x_n^2}$ is the Laplacian, and  $\mu(\tau)$ is an
arbitrary smooth function. Obviously, the cases $n=1,2,3$ are the
most important for prospective real-world applications. In what
follows we assume that
  $\gamma$ is an  arbitrary constant satisfying
the restriction  $\gamma(\gamma+1)\neq0,$  i.e. the equation in
question is nonlinear.

According to the classical Lie method (see, e.g.,   the  recent
monographs \cite{arrigo-15,ch-se-pl-book}), the most general form of
Lie symmetry operators for  (\ref{3-1}) is \[
X=\xi^t(t,\tau,x,u)\partial_t+\xi^{\tau}(t,\tau,x,u)\partial_{\tau}+\xi^i(t,\tau,x,u)\partial_{x_i}+\eta(t,\tau,x,u)\partial_u,
\] where  $\xi^t, \ \xi^{\tau}, \ \xi^i,$ and $ \eta$  are
to-be-determined functions. In the case of the specific function
$\mu(\tau)$, Lie symmetries of (\ref{3-1}) with the fixed number of
the space variables can be easily  calculated using the computer
algebra packages employed in Maple, Mathematica etc. However, we
want to solve {\it the Lie symmetry classification (LSC) problem,
i.e. to identify  all possible forms of the function $\mu(\tau)$ and
$\gamma$  leading to extensions of a so-called principal algebra}.
According to the definition,
 the  principal algebra of equation (\ref{3-1}) is calculated under assumption that $\mu(\tau)$ and $\gamma$ are arbitrary. The detailed algorithm for solving LSC problem for a given PDE involving  arbitrary function(s) as parameter(s) is described
  in \cite[Chapter 2]{ch-se-pl-book}.
 Here we apply this algorithm in the case of the nonlinear PDE (\ref{3-1}).

Now  we present a statement about the group of equivalence
transformations (ETs) of equation (\ref{3-1}) with
$\gamma(\gamma+1)\neq0$. First of all, it can easily be noted that
there is no any local transformation which transforms (\ref{3-1})
to that of the same form but with another parameter
$\gamma^*\not=\gamma$. Thus, ETs can change only the form of the
function $\mu(\tau)$.

\begin{theorem} \label{th0} The group of the continuous ETs   transforming
   equation (\ref{3-1}) with
$\gamma(\gamma+1)\neq0$  to that with the same structure:
\be\label{3-1*} u^*_{t^*}+u^*_{\tau^*}=\Delta
u^*+\left(1-\mu^*(\tau^*)\right)u^*-{\left(u^*\right)}^{\gamma+1},
\ee is an infinite-parameter Lie group generated by the
transformations
 \be\label{a3}\begin{array}{l} t^*=\alpha^2(t+T\left(t-\tau\right)),
 \ \tau^* =\alpha^2(\tau+\tau_0), \\
x^*=\alpha\,\textit{\textbf{R}}\left(t-\tau\right)\,x+X\left(t-\tau\right), \\
u^*=\alpha^{-\frac{2}{\gamma}}u, \
 \mu^*=\frac{\mu-1}{\alpha^2}+1.
 \ea\ee
 Here $\alpha>0$
and $\tau_0$    are the real group parameters,
$x^*=\left(x^*_1,x^*_2,\dots,x^*_n\right)^\top$,
$x=\left(x_1,x_2,\dots,x_n\right)^\top$,
$\textit{\textbf{R}}\left(t-\tau\right)$ is the  $n\times n$
rotation matrix,
$X\left(t-\tau\right)=\left(X^1\left(t-\tau\right),X^2\left(t-\tau\right),\dots,X^n\left(t-\tau\right)\right)^\top$,
and $X^i$ ($i=1,\dots,n$) are arbitrary smooth function. The
function $T$ is an arbitrary smooth function with the restriction
 $T\not=-(t-\tau).$
\end{theorem}

\begin{small}
\begin{table}[h!]
\caption{Lie symmetries of the nonlinear equation
(\ref{3-1})}\medskip
\label{tab1*}       
\begin{tabular}{|c|c|c|}
\hline  & Restrictions  &  Additional Lie symmetries  \\
 \hline && \\
  1 & $\mu(\tau)={\rm const}, \ \mu(\tau)\neq 1$  & $F^2(t-\tau)\p_{\tau}$ \\ \hline &&\\
 2 & $\mu(\tau)=1+\frac{\alpha}{\tau }$  & $F^3(t-\tau)\left(2t\p_t+2\tau\p_{\tau}+x_i\p_{x_i}-\frac{2}{\gamma}\,u\p_u\right)$ \\
 \hline &&\\
 3 & $\mu(\tau)=1$  & $F^2(t-\tau)\p_{\tau}, \
F^3(t-\tau)\left(2t\p_t+2\tau\p_{\tau}+x_i\p_{x_i}-\frac{2}{\gamma}\,u\p_u\right)$ \\
  \hline &&\\
   4 & $\gamma=1, \ \mu(\tau)=1 - 2 \tan \tau$
   &
$F^4(t-\tau)\left(\p_{\tau}+\sec^2\tau\p_u\right)$
 \\
    \hline &&\\
   5 & $\gamma=1, \ \mu(\tau)=1 + 2\tanh \tau$  & $F^5(t-\tau)\left(\p_{\tau}-\operatorname{sech}^2\tau\p_u\right)$ \\
  \hline &&\\
  6 & $\gamma=1, \ \mu(\tau)=1 + 2\coth \tau$  & $F^6(t-\tau)\left(\p_{\tau}+\operatorname{csch}^2\tau\p_u\right)$ \\
  \hline &&\\
   7 & $\gamma=1, \ \mu(\tau)=1+\frac{2}{\tau}$  & $F^3(t-\tau)\left(2t\p_t+2\tau\p_{\tau}+x_i\p_{x_i}-2\,u\p_u\right)$,
 \\
&&  $F^7(t-\tau)\left(\p_{\tau}+\tau^{-2}\p_u\right)$ \\
   \hline
   &&\\
   8 & $\gamma=1, \ \mu(\tau)=1+\frac{1-2\alpha\tan\left(\alpha\ln
\tau\right)}{\tau}$  &
$F^8(t-\tau)\Big(2t\p_t+2\tau\p_{\tau}+x_i\p_{x_i}+$ \\
& &$ \left(\frac{2\alpha^2}{\tau}\sec^2\left(\alpha\ln
\tau\right)-2u\right)\p_u\Big)$ \\
   \hline
   &&\\
   9 & $\gamma=1, \
\mu(\tau)=1+\frac{1+2\alpha\tanh\left(\alpha\ln \tau\right)}{\tau}$
&
$F^9(t-\tau)\Big(2t\p_t+2\tau\p_{\tau}+x_i\p_{x_i}- $ \\
& $\alpha\neq\pm\frac{1}{2}$ &
$\left(\frac{2\alpha^2}{\tau}\operatorname{sech}^2\left(\alpha\ln
\tau\right)+2u\right)\p_u\Big)$\\
   \hline
   &&\\
   10 & $\gamma=1, \
\mu(\tau)=1+\frac{1+2\alpha\coth\left(\alpha\ln \tau\right)}{\tau}$
&
$F^{10}(t-\tau)\Big(2t\p_t+2\tau\p_{\tau}+x_i\p_{x_i}+ $ \\
& $\alpha\neq\pm\frac{1}{2}$ &
$\left(\frac{2\alpha^2}{\tau}\operatorname{csch}^2\left(\alpha\ln
\tau\right)-2u\right)\p_u\Big)$\\
   \hline
   &&\\
   11 & $\gamma=1, \ \mu(\tau)=1+\frac{2+\ln
\tau}{\tau\ln \tau}$  &
$F^{11}(t-\tau)\Big(2t\p_t+2\tau\p_{\tau}+x_i\p_{x_i}+$ \\
& & $\left(\frac{2}{\tau\ln^{2}
\tau}-2u\right)\p_u\Big)$ \\
   \hline
 \end{tabular}
\end{table}
\end{small}

\textbf{Proof of Theorem~\ref{th0}}  is based on the known technique
for constructing the group of continuous ETs. Because this technique
is rather cumbersome, there are not many papers,
 in which one was described in detail and  successfully employed
 for nontrivial PDEs (good examples can be found in  \cite{ibr-tor-1991,lahno-2002}).

 Thus, we should start
from the infinitesimal operator \be\label{2-9}
E=\xi^t(t,\tau,x,u)\partial_t+\xi^{\tau}(t,\tau,x,u)\partial_{\tau}+\xi^i(t,\tau,x,u)\partial_{x_i}+\eta(t,\tau,x,u)\partial_u+
\zeta(t,\tau,x,u,\mu)\partial_{\mu}, \ee where $\xi^t, \ \xi^{\tau},
\ \xi^i,\ \eta$ and $\zeta$  are to-be-determined functions. We note
that formula (\ref{2-9}) without the last term  gives  the most
general form of Lie symmetries  of  equation (\ref{3-1}).

In order to find the operator $E$, we should apply the invariance
criteria \be\label{2-8}\mbox{\raisebox{-1.6ex}{$\stackrel{\displaystyle  
E}{\scriptstyle 2}$}}\,  
\left(\Delta u-u_t-u_{\tau}+(1-\mu)u-u^{\gamma+1}\right)  
\Big\vert_{{\cal{M}}}=0, \ \mbox{\raisebox{-1.6ex}{$\stackrel{\displaystyle  
E}{\scriptstyle 1}$}}\,  
\left(S^k\right)  
\Big\vert_{{\cal{M}}}=0, \ k=1,2,3,\ee where $\mbox{\raisebox{-1.6ex}{$\stackrel{\displaystyle  
E}{\scriptstyle 1}$}}$ and \mbox{\raisebox{-1.6ex}{$\stackrel{\displaystyle  
E}{\scriptstyle 2}$}} represent the first- and second-order
prolongations of the operator (\ref{2-9}), the manifold \[{\cal{M}}
= \{\Delta u-u_t-u_{\tau}+(1-\mu)u-u^{\gamma+1}=0,\ S_1\equiv
\mu_t=0,\ S_2\equiv \mu_x=0,\ S_3\equiv \mu_u=0\}.\]

After the straightforward calculations using formulae (\ref{2-8}),
we arrive at the infinitesimal operator \be\label{2-10}
E=\xi^t(t,\tau)\partial_t+\xi^{\tau}(\tau)\partial_{\tau}+\xi^i(t,\tau,x)\partial_{x_i}+\left(\eta^1(t,\tau,x)u+\eta^0(t,\tau,x)\right)\partial_u+
\zeta(\tau,\mu)\partial_{\mu}, \ee where the coefficients are
to-be-determined functions that  satisfy the  system of linear PDEs
\begin{eqnarray} && \xi^t_t+\xi^t_{\tau}-2\xi^i_{x_i}=0,\label{2-11} \\
&& \xi^i_{x_j}+\xi^j_{x_i}=0, \ i\neq j, \label{2-12} \\
&& \xi^t_t+\xi^t_{\tau}=\xi^{\tau}_{\tau}, \label{2-13}\\
&& 2\eta^1_{x_i}+\xi^i_{t}+\xi^i_{\tau}=0,\label{2-14} \\
&&
\left(\xi^t_t+\xi^t_{\tau}+\gamma\eta^1\right)u^{\gamma+1}+\left(\gamma+1\right)\eta^0u^{\gamma}+\nonumber\\
&& \Big(\eta^1_t+\eta^1_{\tau}-\Delta
\eta^1+\zeta-(1-\mu)\left(\xi^t_t+\xi^t_{\tau}\right)\Big)u+
\eta^0_t+\eta^0_{\tau}-\Delta \eta^0-(1-\mu)\eta^0=0. \label{2-15}
\end{eqnarray}

Consider equation (\ref{2-15}). Since the parameter $\gamma$  is
arbitrary, one immediately obtains
\[\eta^0=0, \ \eta^1=-\frac{\xi^t_t+\xi^t_{\tau}}{\gamma}, \ \zeta=(1-\mu)\xi^{\tau}_{\tau}+\frac{\xi^{\tau}_{\tau\tau}}{\gamma}.\]
Solving the linear equations (\ref{2-11})--(\ref{2-14}), we
determine the coefficients of  the operator (\ref{2-9})\,:
\be\label{2-16}\ba{l} \xi^t=2\beta t+F^1\left(t-\tau\right), \
\xi^{\tau}=2\beta\tau+\beta_0,
\medskip \\
\xi^{i}=G^i\left(t-\tau\right)+\sum\limits_{j=1}^{n}H^{ij}\left(t-\tau\right)x_j, \ i=1,\dots,n, \medskip \\
\eta=-\frac{2\beta}{\gamma}u, \ \zeta=2\beta(1-\mu),\ea\ee where
$\beta_0$ and $\beta$ are arbitrary constants; $F^1$ and $G^i$ are
arbitrary smooth functions. The functions $H$ with superscripts
satisfy the conditions $H^{ij}+H^{ji}=0$ for $i\neq j$,
$H^{ii}=\beta$.

 Finally, using the well-known formulae allowing to
construct the Lie group, which corresponds to  the Lie algebra
generated by  (\ref{2-9}) with the coefficients  (\ref{2-16}), the
Li group (\ref{a3}) has been identified. Because the Lie algebra
obtained is infinite-dimensional, we present some details in
Appendix.

\emph{\textbf{The proof is completed.}}

\begin{theorem} \label{th1} Equation (\ref{3-1}) with an arbitrary parameter $\gamma$ and an arbitrary
function $\mu$   is invariant under the  infinity-dimensional
principal algebra that  is generated by the Lie symmetry  operators:
\be\label{3-3} F^1(t-\tau)\p_t, \ G^i(t-\tau)\p_{x^i}, \
H^{ij}(t-\tau)(x_i\p_{x_j}-x_j\p_{x_i}), \ee where $F^1, \ G^i$ and
$H^{ij}$ are arbitrary smooth functions  of variable $t-\tau$, \
$i=1,\dots,n, \ j=1,\dots,n, \ (i<j).$ In other words, operators
(\ref{3-3}) form the principal algebra of invariance of equation
(\ref{3-1}).
\end{theorem}


\begin{theorem} \label{th2}
 Equation (\ref{3-1}) with
$\gamma(\gamma+1)\neq0$ depending on  $\mu(\tau)$ and $\gamma$
admits  eleven extensions of the principal algebra (\ref{3-3}) that
are listed in Table~\ref{tab1*}. Any other equation of the form
(\ref{3-1}) is either invariant w.r.t. the principal algebra, or is
reducible to one listed in  Table~\ref{tab1*} by the ETs (\ref{a3}).
\end{theorem}

\textbf{Proof of Theorems~\ref{th1} and~\ref{th2}.}

Because the nonlinear equation  (\ref{3-1}) involves only the
single  function
  $\mu$ as a parameter, we use a simplification of the algorithm that
   was  suggested in  \cite[Chapter 2]{ch-se-pl-book}.
   First of all, we note that the group of ETs (\ref{a3}) will be used at the final stage in order to reduce some arbitrary parameters to the values 0 or 1.


At the first step, we construct a system of determining equations.
Applying the infinitesimal criterion of invariance to the most
general form of  Lie symmetries  of  equation (\ref{3-1})  (see
(\ref{2-9}) without the last term), one immediately obtains
\[ \xi^t_{x_i}=\xi^\tau_{x_i}=\xi^t_{u}=\xi^\tau_{u}=\xi^i_u=0, \ i=1,\dots,n, \quad \eta= \eta^1(t,\tau,x)u+\eta^0(t,\tau,x).\]
So
\[
X=\xi^t(t,\tau)\partial_t+\xi^{\tau}(t,\tau)\partial_{\tau}+
\xi^i(t,\tau,x)\partial_{x_i}+\left(\eta^1(t,\tau,x)u+\eta^0(t,\tau,x)\right)\partial_u,
\] where  $\xi^t, \ \xi^{\tau}, \ \xi^i, \ \eta^0$ and $\eta^1$ are
to-be-determined  functions. Using the above form of Lie symmetries,
the system of determining equations was derived:
\begin{eqnarray} && \xi^t_t+\xi^t_{\tau}-2\xi^i_{x_i}=0,\label{3-11} \\
&& \xi^i_{x_j}+\xi^j_{x_i}=0, \ i< j, \label{3-12} \\
&& \xi^t_t+\xi^t_{\tau}=\xi^{\tau}_t+\xi^{\tau}_{\tau}, \label{3-12*} \\
&& 2\eta^1_{x_i}+\xi^i_{t}+\xi^i_{\tau}=0,\label{3-13} \\
&&
\left(\xi^t_t+\xi^t_{\tau}+\gamma\eta^1\right)u^{\gamma+1}+\left(\gamma+1\right)\eta^0u^{\gamma}+\nonumber\\
&& \Big(\eta^1_t+\eta^1_{\tau}-\Delta
\eta^1+\mu'\,\xi^{\tau}-(1-\mu)\left(\xi^t_t+\xi^t_{\tau}\right)\Big)u+
\eta^0_t+\eta^0_{\tau}-\Delta \eta^0-(1-\mu)\eta^0=0. \label{3-14}
\end{eqnarray}
We remind the reader that  $i,j= 1,\dots,n$.

At the second step, the parameters $\gamma$ and $\mu$  in equation
(\ref{3-1}) are assumed to be arbitrary.  In fact, according to the
definition,  the principal algebra of invariance consists of all Lie
symmetries, which do not depend on the form of $\gamma$ and $\mu$.
In this case, the result presented in Theorem~\ref{th1} follows
straightforwardly. Indeed,
we immediately obtain  from equation (\ref{3-14}) the following
conditions:
 \be\label{3-15}\eta^1=0, \ \eta^0=0, \
\xi^{\tau}=0, \ \xi^t_t+\xi^t_{\tau}=0. \ee Solving the linear
equations (\ref{3-11})--(\ref{3-13}) and the last one from
(\ref{3-15}), we  derive exactly the coefficients of  the Lie
symmetries  (\ref{3-3}). Thus, the principal algebra of invariance
is identified.

At the third step, we need to find all possible specifications  of
parameters $\gamma$ and $\mu$ that can  lead to inequivalent
extensions of the  principal algebra (\ref{3-3}).  Firstly,  we
split equation (\ref{3-14}) with respect to the different exponents
 $1, \ \gamma $ and $ \gamma+1$ of $u$. Taking into
account that $\gamma(\gamma+1)\neq0,$ we conclude that $\gamma=1$ is
the only special case. Thus, one needs to consider two cases:
\emph{\textbf{1)}} $\gamma$ is an arbitrary parameter and
\emph{\textbf{2)}} $\gamma=1$.

Let us examine Case \emph{\textbf{1)}} in  detail. Splitting
equation (\ref{3-14}) with respect to the above exponents, we obtain
 $\eta^0=0$ and  the  equations \begin{eqnarray} &&
\xi^t_t+\xi^t_{\tau}+\gamma\eta^1=0, \label{3-16} \\
&& \eta^1_t+\eta^1_{\tau}-\Delta
\eta^1+\mu'\,\xi^{\tau}-(1-\mu)\left(\xi^t_t+\xi^t_{\tau}\right)=0.
\label{3-17}
\end{eqnarray}
Equation (\ref{3-16}) immediately gives $\eta^1_{x_i}=0$. So,
solving equations (\ref{3-11})--(\ref{3-13}), we identify the
functions $\xi^t, \ \xi^{\tau}$ and $\xi^i$ as follows:
\[\ba{l}
\xi^t=2t\,F^3\left(t-\tau\right)+F^1\left(t-\tau\right), \
\xi^{\tau}=2\tau\,F^3\left(t-\tau\right)+F^2\left(t-\tau\right),
\medskip \\
\xi^{i}=G^i\left(t-\tau\right)+H^{ij}\left(t-\tau\right)x_j,\ea\]
where $H^{ij}+H^{ji}=0$ for $i< j$. Here, the functions $F^k
(k=1,2,3)$ and $ G^i$
 are arbitrary smooth functions. Substituting the functions $\xi^t$ and $\xi^{\tau}$
 derived above into (\ref{3-16}), we find
 the function
$\eta^1=-\frac{2}{\gamma}\ F^3.$

Thus, to complete the examination  of Case \emph{\textbf{1)}}, one
needs to solve the classification equation (\ref{3-17}) that now can
be rewritten in the following form \be\label{3-19}
\mu'\left(2\tau\,F^3+F^2\right)=2(1-\mu)F^3.\ee To do this, one
needs  to consider two different subcases: $\mu'F^3=0$ and
$\mu'F^3\neq0$.

The first subcase leads to the following results:  Case 1 of
Table~\ref{tab1*} if $\mu'=0$ and $\mu\neq1$; Case 3 of
Table~\ref{tab1*}  if $\mu=1$. If  $\mu'\neq0$ then only the
principal algebra (\ref{3-3}) is obtained.

In the second subcase,  equation (\ref{3-19}) can be  rewritten  in
the form \be\label{3-20}
\frac{F^2}{2F^3}=\frac{1-\mu}{\mu'}-\tau.\ee Since $\mu$  does not
depend on the variable $t$, equation (\ref{3-20}) leads to the
linear ODE
\[ \frac{1-\mu}{\mu'}-\tau=\tau_0. \]
 Solving the above equation, one obtains
 \[ \mu= 1+ \frac{\alpha}{\tau +\tau_0}, \ \alpha\not=0, \]
 where $\tau_0$ and $\alpha$ is an arbitrary constants.
 Using the ET $\tau +\tau_0 \to \tau$ (see the first line in (\ref{a3}))
 the above functions reduces to the form $\mu= 1+ \frac{\alpha}{\tau}$.
 Moreover, it means that $F^2=0$ (see (\ref{3-20})), while the function $F^3$ remains arbitrary (refer to formula (\ref{3-19})).
 As a result, the additional operator
 \[ F^3(t-\tau)\left(2t\p_t+2\tau\p_{\tau}+x_i\p_{x_i}-\frac{2}{\gamma}\,u\p_u\right) \]
 is derived. Thus,
  Case 2 of Table~\ref{tab1*} is identified.

In Case \emph{\textbf{2)}},
a quite similar  analysis leads to the results presented in Cases
4--11 of Table~\ref{tab1*}, in which $\alpha\not=0$ and $\beta$ are
arbitrary constants.

Notably, the group of ETs (\ref{a3}) (more complicated
transformations then the  $\tau$-translation mentioned above) was
used for simplifications of the function $\mu(\tau)$ in Cases 4--6
and 8--11 of Table~\ref{tab1*}. In Cases 4--6,  the ET
\be\label{3-20*} t^*=\frac{|\alpha|}{2}\,t, \ \tau^*
=\frac{|\alpha|}{2}\,\tau, \ x^*=\sqrt{\frac{|\alpha|}{2}}\,x, \
u^*=\frac{2}{|\alpha|}\,u \ee were used in order to remove  the
parameter $\alpha$, which naturally arises in the expressions
obtained for function $\mu(\tau)$, namely:  $\mu(\tau)=1 - \alpha
\tan\frac{\alpha \tau}{2}$ (Case 4), $\mu(\tau)=1 + \alpha
\tanh\frac{\alpha \tau}{2}$ (Case 5) and $\mu(\tau)=1 + \alpha
\coth\frac{\alpha \tau}{2}$ (Case 6).
In Cases 8--10 and Case 11, the ETs
\[t^*=e^{\frac{\beta}{\alpha}}\,t, \ \tau^* =e^{\frac{\beta}{\alpha}}\,\tau, \
x^*=e^{\frac{\beta}{2\alpha}}\,x, \
u^*=e^{-\frac{\beta}{\alpha}}\,u\] and
\[t^*=e^{\beta}\,t, \ \tau^* =e^{\beta}\,\tau, \
x^*=e^{\frac{\beta}{2}}\,x, \ u^*=e^{-\beta}\,u, \] removing the
parameter $\beta$,
 where used for the same purposes.
As a result, Table~\ref{tab1*} is derived.

\emph{\textbf{The proof is completed.}}


\begin{remark}
The nonlinear equation (\ref{3-1})  is reduced to equation
\be\label{3-1**} u_{\tau^*}=\Delta
u+\big(1-\mu(\tau^*)\big)u-u^{\gamma+1}\ee by the substitution
\[t^*=t-\tau, \ \tau^*=\tau,\] in which the function $u$ should be
still considered as function on $n+2$ independent variables. The Lie
symmetry  analysis is a little bit simpler for (\ref{3-1**}).
However,  we would still go back to the original variables because
the variable $t^*=\tau-t$ does not have any biological meaning in
contrast to $\tau $ and $t$. Taking into account that one of the
main goals of this work  is application of the results for modelling
of epidemics, we prefer to keep biologically motivated variables.
\end{remark}

In conclusion of this section, we present the following observation.
Equation (\ref{3-1}) with $\gamma=1$, i.e. \be\label{3-4}
u_t+u_{\tau}=\Delta u+(1-\mu(\tau))u-u^{2} \ee can be rewritten in
the form \[ u_t+u_{\tau}=\Delta
u-u^{2}+\frac{\mu'}{2}+\frac{(1-\mu)^2}{4}\] by using the
substitution \be\label{3-6} u \ \rightarrow \ u+\frac{1-\mu}{2}.\ee

In particular,  equation (\ref{3-4}) with $\mu=1+\frac{2}{\tau}$
(see Case 7 in Table~\ref{tab1*}) is reduced to the equation
\[u_t+u_{\tau}=\Delta u-u^2\]
by the transformation \be\label{3-6*} u \ \rightarrow \
u-\frac{1}{\tau}.\ee Now one notes that the above equation
corresponds  to Case 3 of Table~\ref{tab1*}. Obviously,
(\ref{3-6*}) is not a ET, therefore it  is a form-preserving
transformation (FPT).
  We note that FPTs (equivalent  terminology is `admissible transformations' \cite{ga-wint-92})
  were introduced in
 \cite{kingston-91,ki-sopho-93} for study nonlinear PDEs. Later it was shown that FPTs
 play essential role in solving  LSC of PDEs belonging to a given class
 (see \cite[Chapter 2]{ch-se-pl-book} and papers cited therein).
Notably,  a sophisticated approach for so-called enhanced group
analysis based
 on admissible transformations  was
 independently  developed in \cite{van-pop-sopho-09}.
 A further Lie symmetry classification of (\ref{3-1}) based on
 form-preserving/admissible transformations
  lies beyond the specific scopes of this study.

\section{Exact solutions} \label{sec-4}

Let us construct exact solutions of Eq. (\ref{3-1}) in the case
$n=1$: \be\label{4-1}
u_t+u_{\tau}=u_{xx}+(1-\mu(\tau))u-u^{\gamma+1}.\ee It can be noted
that the nonlinear equation (\ref{4-1}) is reduced to equation
\be\label{4-2}
u_{t^*}=u_{xx}+\big(1-\mu(t^*-\tau^*)\big)u-u^{\gamma+1}\ee by the
substitution \be\label{4-3}t^*=\frac{t+\tau}{2}, \
\tau^*=\frac{t-\tau}{2}.\ee Equation (\ref{4-2}) is nothing else but
a  nonlinear reaction-diffusion (RD) equation involving the variable
$\tau^*$ as a  parameter in  the reaction term.

\subsection{Travelling wave type solutions}

Firstly, we examine
equation (\ref{4-2}) with a constant function  $\mu(\tau)$, i.e. the
RD equation
\be\label{4-3*} u_{t^*}=u_{xx}+ \lambda u-u^{\gamma+1},\ee where
$\lambda=1-\mu ={\rm const}$. Exact solutions of equation
(\ref{4-3*}) with arbitrary and  correctly-specified $\gamma$ were
found in several studies. In particular, known  travelling waves
(TWs) are summarized in \cite{gild-ker-04}. So, taking a known  TW
solution and replacing arbitrary constants by arbitrary functions of
the variable  $\tau^*=\frac{t-\tau}{2}$, one readily obtains an
exact solution of equation  (\ref{4-1}) with $\mu(\tau)=1-\lambda$.
 In particular, using the travelling wave  presented in \cite{ch-dav-2022-cd} (see (3.6)
therein)
 one arrives at the following  solution of equation (\ref{4-1})
\be\label{4-5} u(t,\tau,x) =\lambda^{1/\gamma}\left(1+
C\left(t-\tau\right)\exp
\left(\pm\frac{\gamma\sqrt{\lambda}}{\sqrt{2(\gamma+2)}}\,x-\frac{\lambda\gamma(\gamma+4)}{4(\gamma+2)}\,(t+\tau)\right)\right)^{-2/\gamma},\ee
  where $\lambda=1-\mu>0$ and
$C(t-\tau)$ is an arbitrary smooth function. Obviously, (\ref{4-5})
with $C={\rm const}>0$ and $\gamma>0$ is a typical TW.

Equation (\ref{4-3*}) with  $\gamma=2$:
\[ u_{t^*}=u_{xx}+ \lambda u-u^{3},\] possesses more
complicated solutions that were constructed for the first time in
\cite{kawa-tana-83} and  later were identified  using
$Q$-conditional (nonclassical) symmetries in
 \cite{cla-mans-94, a-h-b-94} (see also \cite[Chapter 4]{ch-se-pl-book}). Assuming $\lambda>0$, the relevant solution has the form
 \[
 u =
 \frac{ \sqrt{\lbd}\,C_2\exp\Bigr[\frac{\sqrt{2\lbd}}{2}x+\frac{3\lbd}{2}t^*\Bigr] -\sqrt{\lbd}\,C_3\exp\Bigr[-\frac{\sqrt{2\lbd}}{2}x+\frac{3\lbd}{2}t^*\Bigr]}
 {C_1+C_2\exp\Bigr[\frac{\sqrt{2\lbd}}{2}x+\frac{3\lbd}{2}t^*\Bigr] +C_3\exp\Bigr[-\frac{\sqrt{2\lbd}}{2}x+\frac{3\lbd}{2}t^*\Bigr]}.
 \]
 Thus,  equation (\ref{4-1}) with  $\mu(\tau)=1-\lambda$ and $\gamma=2$ possesses the following  family of exact solutions involving three arbitrary functions:
\be\label{4-ad3}
 u =
 \frac{ \sqrt{\lbd}\,C_2\left(t-\tau\right)\exp\Bigr[\frac{\sqrt{2\lbd}}{2}x+\frac{3\lbd}{4}(t+\tau)\Bigr]-
 \sqrt{\lbd}\,C_3\left(t-\tau\right)\exp\Bigr[-\frac{\sqrt{2\lbd}}{2}x+\frac{3\lbd}{4}(t+\tau)\Bigr]}
 {C_1\left(t-\tau\right)+C_2\left(t-\tau\right)\exp\Bigr[\frac{\sqrt{2\lbd}}{2}x+\frac{3\lbd}{4}(t+\tau)\Bigr] +
 C_3\left(t-\tau\right)\exp\Bigr[-\frac{\sqrt{2\lbd}}{2}x+\frac{3\lbd}{4}(t+\tau)\Bigr]}.  \ee
 In the special case, stationary (time-independent) solutions of the form
 \be\label{4-ad4}
 u =
 \frac{ \sqrt{\lbd}\,c_2\exp\Bigr[\frac{\sqrt{2\lbd}}{2}x+\frac{3\lbd}{2}\tau\Bigr] -\sqrt{\lbd}\,c_3\exp\Bigr[-\frac{\sqrt{2\lbd}}{2}x+
 \frac{3\lbd}{2}\tau\Bigr]}
 {c_1+c_2\exp\Bigr[\frac{\sqrt{2\lbd}}{2}x+\frac{3\lbd}{2}\tau\Bigr] +c_3\exp\Bigr[-\frac{\sqrt{2\lbd}}{2}x+\frac{3\lbd}{2}\tau\Bigr]}  \ee
  can be identified from the above family of solutions by choosing the functions $C_1=c_1, \ C_i=c_i \exp\Bigr(\frac{3\lbd}{4}(\tau-t)\Bigr), \
   i=2,3 $ ($c_i$ are arbitrary constants).
   Interestingly, the exact solution (\ref{4-ad4}) with $c_3=0$ and $c_1c_2>0$ is a  TW if one considers
   $\tau$ as an analog of time (the same occurs in the case $c_2=0$ and $c_1c_3>0$).

\begin{figure}[t]
\begin{center}
\includegraphics[width=7cm]{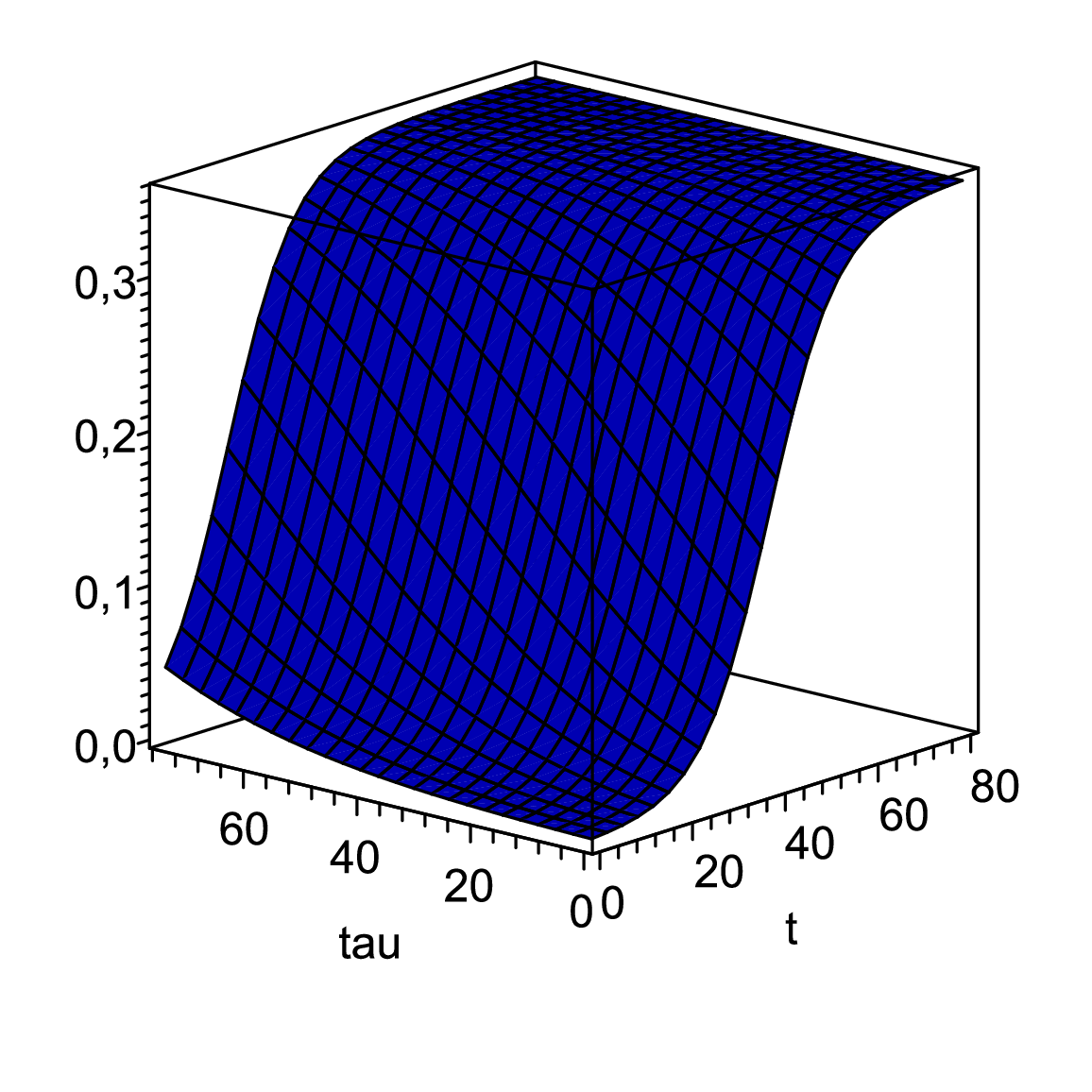}
\caption{3D plot of the  function $U(t,\tau)$ using formula
(\ref{4-23})  with
 $x_0=1, \ \lambda=1/8, \ c_1=25, \ c_2=1/10$.} \label{fig-1}
\end{center}
\end{figure}

\textbf{Example.} Let us examine the properties of the exact
solution (\ref{4-ad3}) in the domain $G=\left\{ (t,\tau,x) \in (0,+
\infty )\times (0,\tau_{\max})\times (0,x_0)\rg\}$. The total number
of infected individuals of the age $\tau $  up to time $t$ is
calculated by the formula
\[U(t,\tau)=\int_0^{x_0}u(t,x,\tau)dx=\sqrt{2}\,
\ln\left(\frac{C_2\,e^{\frac{\sqrt{2\lbd}}{2}\,x_0}+C_3\,e^{-\frac{\sqrt{2\lbd}}{2}\,x_0}+
C_1e^{-\frac{3\lambda(t+\tau)}{4}}}{C_2+C_3+C_1e^{-\frac{3\lambda(t+\tau)}{4}}}\right).
\]
In order to provide meaningful interpretation for modelling of an
epidemic, one requires that  the function $U(t,\tau)$ is
nonnegative, nondecreasing and bounded for all $t>0$ and
$\tau\in(0,\tau_{\max})$. These requirements  can be easily
satisfied  taking into account arbitrariness  of  $C_1, \ C_2$ and
$C_3$. Setting, for example,
$C_1=c_1e^{-\frac{3\lambda(t-\tau)}{8}},\ C_2=c_2$ and $C_3=0$
($c_1>0$ and $c_2>0$ are arbitrary constants), the function
$U(t,\tau)$ takes the form \be\label{4-23}U(t,\tau)=\sqrt{2}\,
\ln\left(\frac{c_2\,e^{\frac{\sqrt{2\lbd}}{2}\,x_0}+
c_1e^{-\frac{3\lambda(3t+\tau)}{8}}}{c_2+c_1e^{-\frac{3\lambda(3t+\tau)}{8}}}\right).
\ee This function is nonnegative, nondecreasing, and bounded for any
positive parameters $\lambda, \ x_0, \ c_1$ and $c_2$. An example of
the $U(t,\tau)$ plot of
  is presented in Fig.~\ref{fig-1} (we remind the reader that equation (\ref{4-1}) is presented in
  the nondimensional form, see formulae (\ref{2-5})). It can be easily seen that the function $U(t,\tau)$ is increasing with time and this is in agreement with its meaning, i.e. the total number of the infected individuals
 of the age $\tau$  should grow with time. Moreover, the total number  is growing with the age $\tau$ as well because  typically older individuals are more  affected by epidemics than younger.

\begin{figure}[t]
\begin{center}
\includegraphics[width=7cm]{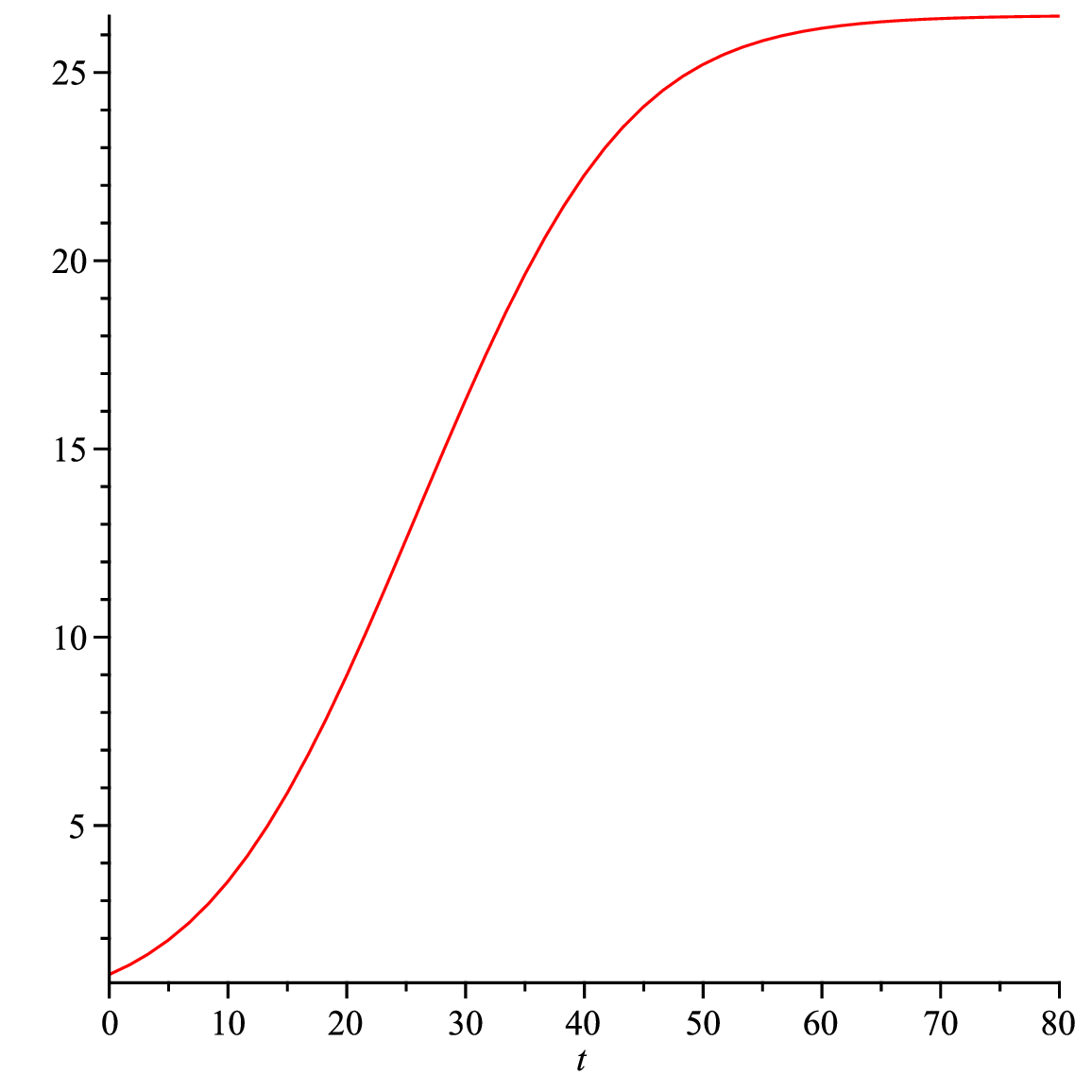}
\caption{Plot of  the  function $U_{\rm total}(t)$ using formula
(\ref{4-24})  with
 $x_0=1, \ \tau_{\max}=75, \ \lambda=1/8,  c_1=25,  c_2=1/10$.} \label{fig-2}
\end{center}
\end{figure}

The total number of infected individuals up to time $t$ is
calculated using
  formula (\ref{2-4*}), hence we obtain\,: \be\label{4-24} U_{\rm total}(t)=\sqrt{2}\int_0^{\tau_{\max}}
\ln\left(\frac{c_2\,e^{\frac{\sqrt{2\lbd}}{2}\,x_0}+
c_1e^{-\frac{3\lambda(3t+\tau)}{8}}}{c_2+c_1e^{-\frac{3\lambda(3t+\tau)}{8}}}\right)d\tau.
\ee The integral arising in (\ref{4-24}) can be expressed in the
terms of the special function dilogarithm, however the  formula
obtained is rather cumbersome and is omitted here. The relevant plot
was drawn
and  is presented in Fig.~\ref{fig-2}. Interestingly, the curve
presented  in Fig.~\ref{fig-2} has the form  that is typical  for
modelling of the total number of infected individuals  during each
wave of  COVID-19 pandemic (see, e.g., examples in
\cite{ch-dav-2020}). Notably such curve is often called sigmoid. It
should be stressed that  the  curves obtained  from    official data
  in many countries qualitatively have the form of sigmoid as well.

 \subsection{Lie solutions }

Now we construct exact solutions of equation  (\ref{4-1}) using Lie
symmetries presented in Theorem~\ref{th1}. Depending on the function
$\mu$, there are several cases when equation  (\ref{4-1})  admits
additional Lie symmetries (see Table~\ref{tab1*}).   Because  this
function means the death rate and  should satisfy natural
restrictions, in particular $\mu>0$  and $ \frac{d\mu}{d\tau}>0$,
let us   consider Case~5 of Table~\ref{tab1*}.  From the
applicability point of view, one expects that $\mu$ involves a
parameter because the death rate cannot be fixed, therefore we
introduce $\alpha>0$ using  ET  (\ref{3-20*}).
 Thus,  our aim is  to solve the equation
\be\label{4-6} u_t+u_{\tau}=u_{xx}-\alpha \tanh\frac{\alpha
\tau}{2}\,u-u^{2}\ee using its Lie symmetry \be\label{4-7}
X=F^1(t-\tau)\p_t+F^5(t-\tau)\p_{\tau}+
G(t-\tau)\p_{x}-\frac{\alpha^2F^5(t-\tau)}{4}\,\operatorname{sech}^2\frac{\alpha
\tau}{2}\p_u.\ee Applying transformation (\ref{4-3}) to
(\ref{4-6})--(\ref{4-7}), we arrive at the equation \be\label{4-8}
u_{t^*}=u_{xx}-\alpha \tanh\frac{\alpha
(t^*-\tau^*)}{2}\,u-u^{2}.\ee Thus, according to Case 5 of Table
~\ref{tab1*} (taking into account  ET  (\ref{3-20*})!)  the most
general form of Lie symmetry of the above equation reads as
\be\label{4-9} X^*=\frac{F^1(\tau^*)+F^5(\tau^*)}{2}\,\p_{t^*}+
\frac{F^1(\tau^*)-F^5(\tau^*)}{2}\,\p_{\tau^*}+ G(\tau^*)\p_{x}-
\frac{\alpha^2}{4} F^5(\tau^*)\,\operatorname{sech}^2\frac{\alpha
(t^*-\tau^*)}{2}\p_u,\ee where $F^1, \ F^5$ and $G$ are arbitrary
functions. Different reductions of (\ref{4-8})  to ODEs  can be
derived depending on these functions. A complete list of the
reductions and their analyses lies beyond the scopes of this study.
Here we present an interesting particular case.

Let us consider a particular case of (\ref{4-9}), setting
$F^1=F^5=1, \ G=g={\rm const}$, that is
 \be\label{4-10} X^*=\p_{t^*}+
g\p_{x}-\frac{\alpha^2}{4}\,\operatorname{sech}^2\frac{\alpha
(t^*-\tau^*)}{2}\p_u.\ee So, the ansatz generated by  the Lie
symmetry operator (\ref{4-10})  is
\be\label{4-11}u=\Phi(\omega_1,\omega_2)-\frac{\alpha}{2}\tanh\frac{\alpha
(t^*-\tau^*)}{2}, \ \omega_1=\tau^*, \ \omega_2=x-gt^*.\ee
Substituting ansatz (\ref{4-11}) into equation (\ref{4-8}), we
obtain the second-order ODE  \[
\Phi_{\omega_2\omega_2}+g\,\Phi_{\omega_2}-\Phi^{2}+\frac{\alpha^2}{4}=0,\]
which takes  the form \be\label{4-17}
\Phi_{\omega_2\omega_2}+g\,\Phi_{\omega_2}+\Phi(\alpha-\Phi)=0,\ee
by applying the substitution \be\label{4-18} \Phi \rightarrow
\Phi-\frac{\alpha}{2}.\ee

Equation (\ref{4-17}) corresponds to the known ODE that arises
when one seeks for TWs  of  the famous Fisher equation
\[u_{t}=u_{xx}+u(\alpha-u),\] using the ansatz $u=\Phi(\omega_2), \
\omega_2=x-gt.$ The  well-known   exact solution of equation
(\ref{4-17}) has the form \be\label{4-19} \Phi=
\frac{\beta^2}{\left(1+C_1\left(\omega_1\right)\,e^{\frac{\beta}{\sqrt{6}}\,\omega_2}\right)^2},
\ee in the case $\alpha=\beta^2, \ g=\frac{5\beta}{\sqrt{6}}$, and
the form \be\label{4-19**} \Phi=
\frac{\beta^2}{\left(1+C_1\left(\omega_1\right)\,e^{\frac{\beta}{\sqrt{6}}\,\omega_2}\right)^2}-\beta^2,
\ee in the case $\alpha=-\beta^2, \ g=\frac{5\beta}{\sqrt{6}}$. Here
$C_1\left(\omega_1\right)$ is an arbitrary smooth function. In
particular, setting $C_1=1$ in (\ref{4-19}), one obtains the
well-known TW of the Fisher equation that was identified for the
first time in \cite{abl-zep}.

Thus, taking into account (\ref{4-3}), (\ref{4-11}), (\ref{4-18}),
and (\ref{4-19})--(\ref{4-19**}) and applying the translation $t \
\rightarrow \ t+\tau_0, \ \tau \ \rightarrow \tau -\tau_0$, we
arrive at the exact solution \be\label{4-19*} u
=\beta^2\left[1+C_1\left(t-\tau\right)\,\exp\left(\frac{\beta}{\sqrt{6}}\,\left(x-\frac{5\beta}{\sqrt{6}}\frac{t+\tau}{2}\right)\right)\right]^{-2}
-\frac{\beta^2}{2}\tanh\frac{\beta^2 }{2}\,(\tau-
\tau_0)-\frac{\beta^2}{2} \ee of equation \[
u_t+u_{\tau}=u_{xx}-\beta^2 \tanh\frac{\beta^2 (\tau
-\tau_0)}{2}\,u-u^{2},\] where $\beta$ and $\tau_0$ are arbitrary
constants.

\begin{figure}[t]
\begin{center}
\includegraphics[width=8cm]{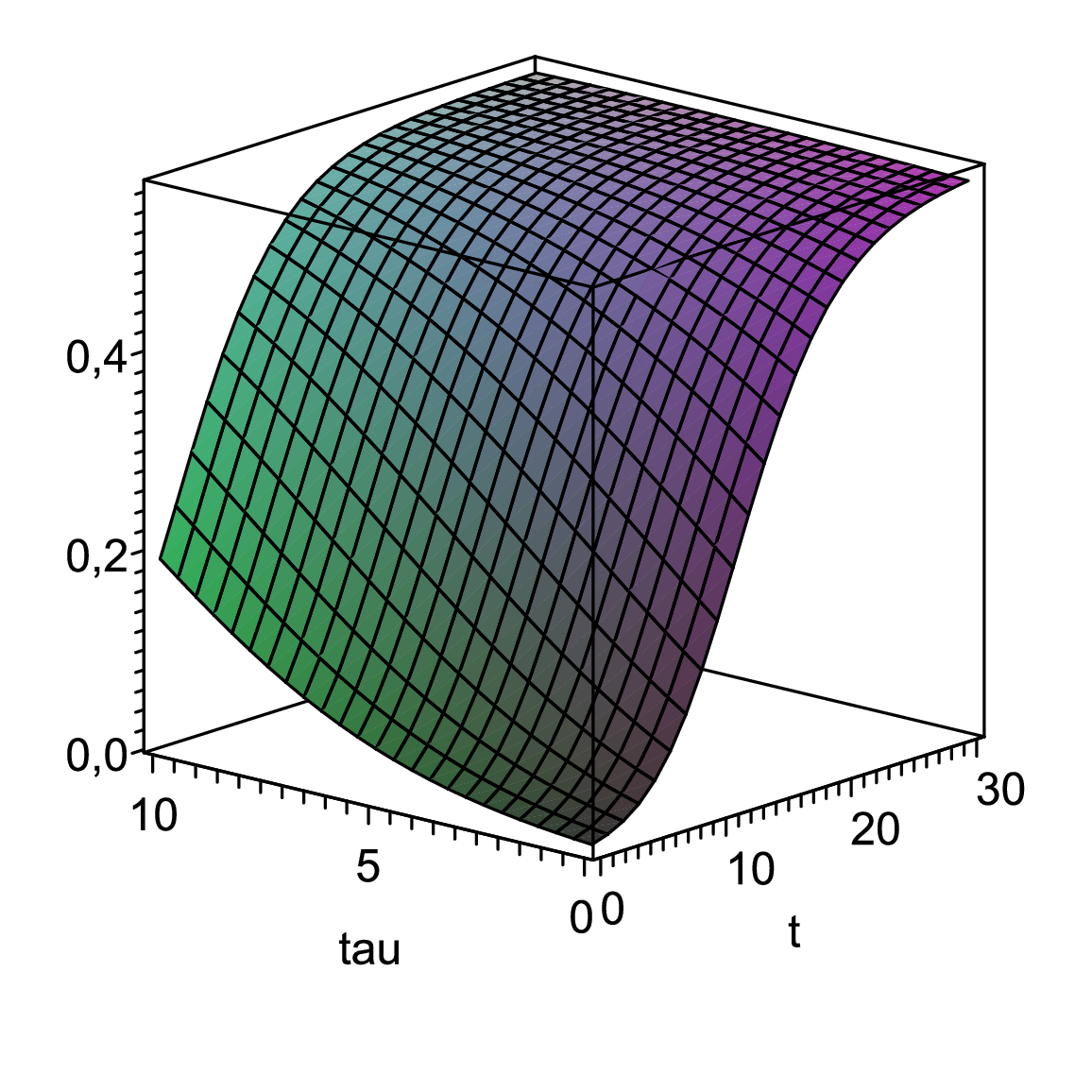}
\includegraphics[width=8cm]{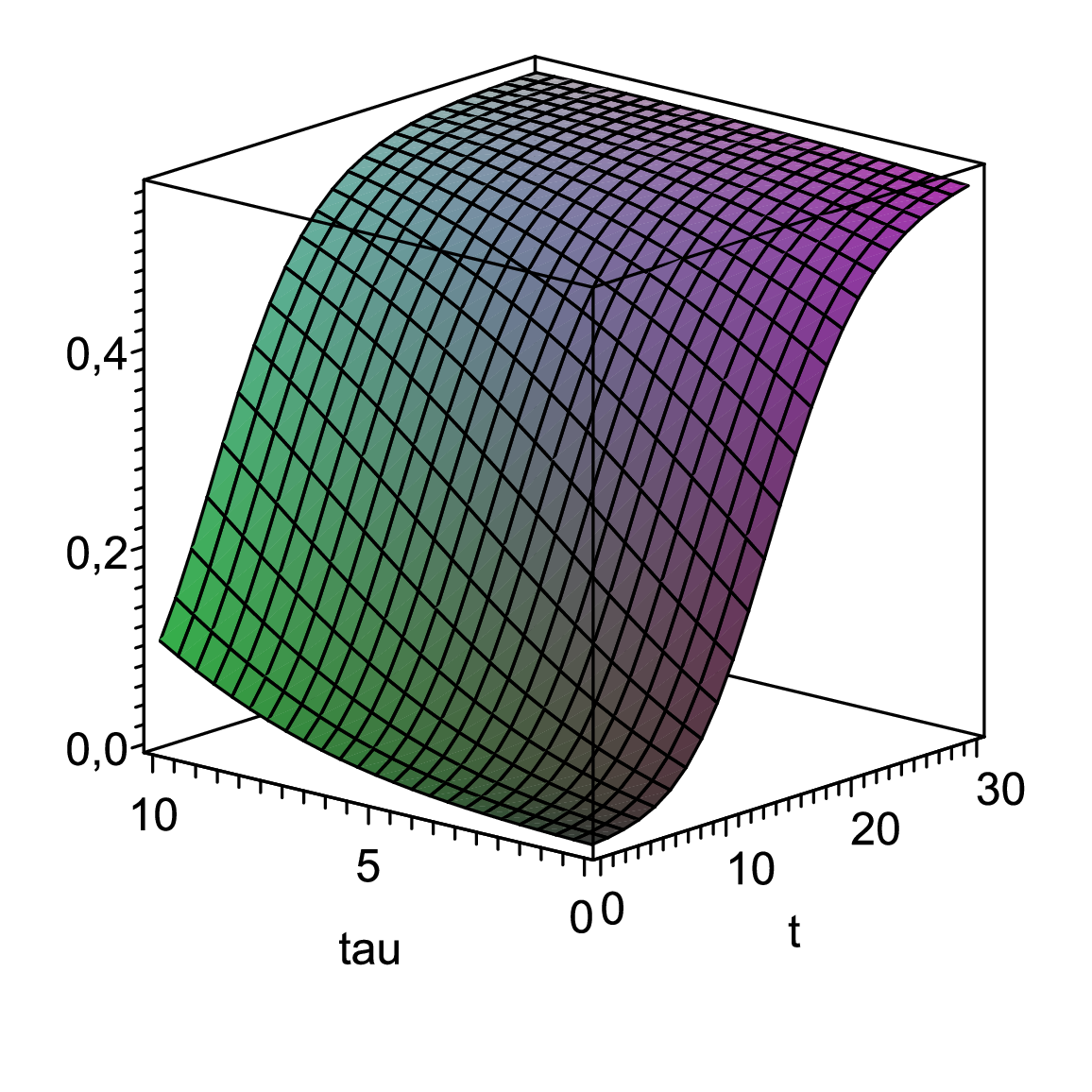}
\caption{3D plots of the  exacty solution  $u(t, \tau, x)$
(\ref{4-19*}) for the fixed points $x=-1$ (left) and $x=1$ (right).
The parameters are fixed as follows:
 $ \beta=3/4, \ C_1=10, \ \tau_0=20$.} \label{fig-3}
\end{center}
\end{figure}

\begin{figure}[t]
\begin{center}
\includegraphics[width=7cm]{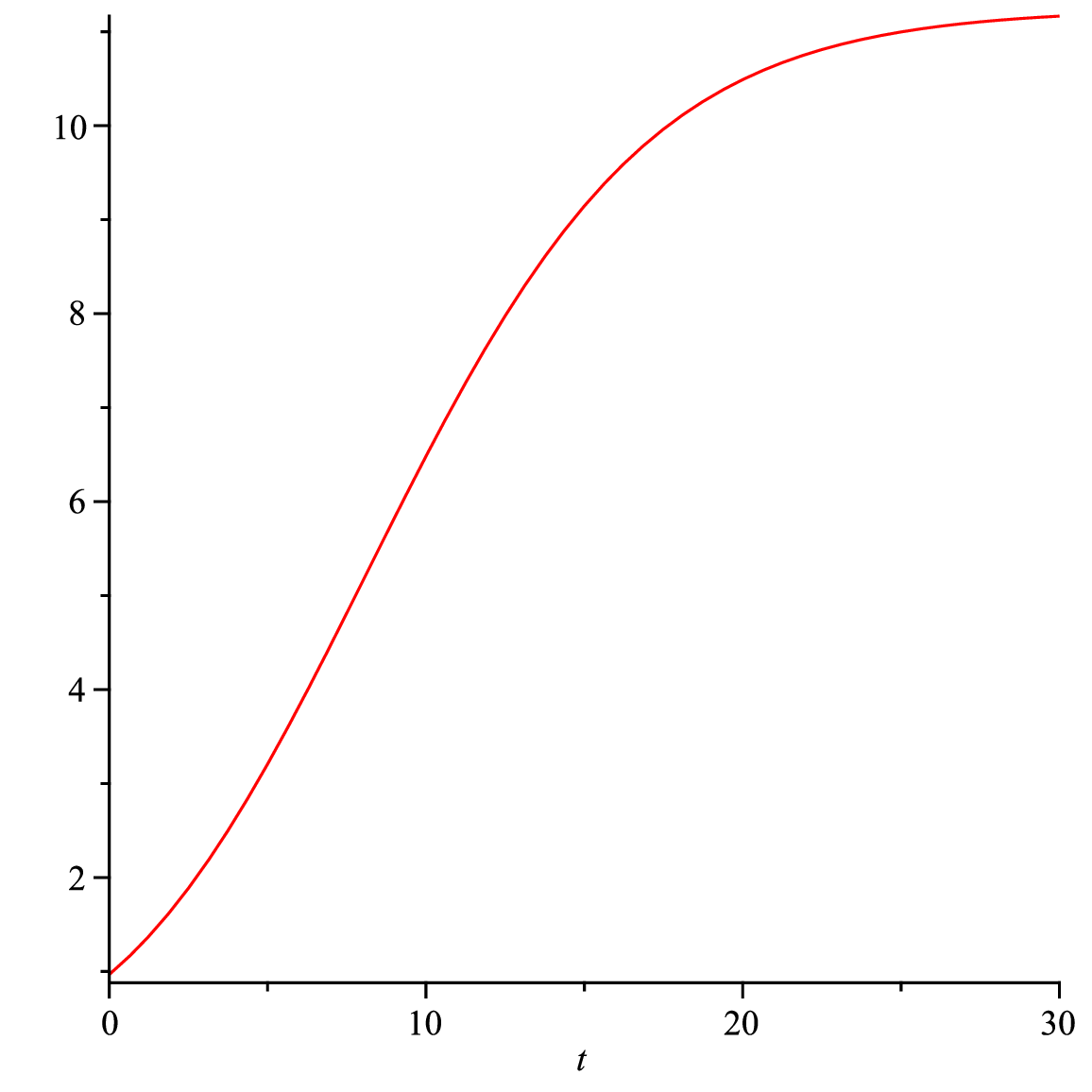}
\caption{Plot of  the  function $U_{\rm total}(t)$ (see formula
(\ref{2-4*})  with $I=[0,10]$ and $\Omega= [-1,1]$) derived using
the exact solution (\ref{4-19*}) with
 $ \beta=3/4, \ C_1=10, \ \tau_0=20$.} \label{fig-4}
\end{center}
\end{figure}

Examples of solution (\ref{4-19*}) are presented in
Fig.~\ref{fig-3}. One may conclude that the exact solution is an
increasing function of the time $t$ and the age $\tau$. Such
behaviour is in agreement with the biological interpretation because
$u$ means the total  density of infected individuals  during the
epidemic spread. In  Fig.~\ref{fig-4}, the curve illustrating the
total number  of infected population is drawn. The function $U_{\rm
total}(t)$ was build using  formula (\ref{2-4*}) with $I=[0,10]$ and
$\Omega= [-1,1]$. As one may easily note, the curve is again  a
sigmoid.

To the best of our knowledge,  the general solution of equation
(\ref{4-17})  is unknown provided   parameters $g$ and $\alpha$ are
arbitrary. In the case $g=\frac{5\beta}{\sqrt{6}}$, the general
solution of this equation has the form: \be\label{4-21}\Phi= \left\{
\begin{array}{l}
6\,e^{-\frac{2\beta}{\sqrt{6}}\,\omega_2}\wp\left(-\frac{\sqrt{6}}{\beta}\,e^{-\frac{\beta}{\sqrt{6}}\,\omega_2}
+C_1(\omega_1);0,C_2(\omega_1)\right), \quad \alpha=\beta^2, \\
6\,e^{-\frac{2\beta}{\sqrt{6}}\,\omega_2}\wp\left(-\frac{\sqrt{6}}{\beta}\,e^{-\frac{\beta}{\sqrt{6}}\,\omega_2}
+C_1(\omega_1);0,C_2(\omega_1)\right)-\beta^2, \quad
\alpha=-\beta^2,
\end{array} \right.\ee
where $\wp$ is the Weierstrass function, $C_1$ and $C_2$ are
arbitrary smooth functions.

In the case  $g=0$,   the general solution of equation (\ref{4-17})
is known  in the form\,: \[
\Phi=6\wp\left(\omega_2+C_1(\omega_1);\frac{\alpha^2}{12},C_2(\omega_1)\right)+\frac{\alpha}{2},\]
therefore the exact solution \[
u(t^*,\tau^*,x)=6\wp\left(x+C_1\left(\tau^*\right);\frac{\alpha^2}{12},C_2\left(\tau^*\right)\right)-\frac{\alpha}{2}\tanh\frac{\alpha
(t^*-\tau^*)}{2}\] of equation (\ref{4-8})  identified.

Finally, applying  substitution (\ref{4-3}), one obtains the
solution \be\label{4-15} u(t,\tau,x)
=6\wp\left(x+C_1\left(t-\tau\right);\, \frac{\alpha^2}{12},
C_2\left(t-\tau\right)\right)-\frac{\alpha}{2}\tanh\frac{\alpha
\,\tau}{2}\ee of the initial  equation (\ref{4-6}). Notably, setting
$C_2=\frac{\alpha^3}{216}$, solution (\ref{4-15}) can be expressed
in terms of  elementary functions:
\[ u(t,\tau,x)
=\frac{\alpha}{2}\left(-1+3\csc^2\left[\frac{\sqrt{\alpha}}{2}\,x+C\left(t-\tau\right)\right]-\tanh\frac{\alpha
\,\tau}{2}\right),\] where $C$ is an arbitrary smooth function.

\begin{remark} All exact solutions presented in this subsection
can be identified from the known solutions of the Fisher equation by
a chain of substitutions. In fact, equation (\ref{4-6}) is related
with the Fisher equation via substitution
$u^*=u+\alpha-\frac{\alpha}{1+e^{\alpha\tau}}$ and (\ref{4-3}).
\end{remark}

\section{Conclusions} \label{sec-5}

In this study, the age-structured diffusive  model with the
governing equation (\ref{2-3}) is suggested for the   mathematical
modelling of epidemics. The model is a generalization of the known
age-structured model based on the linear equation (\ref{2-7}). To
construct the model, the recently developed model
\cite{ch-dav-2022-cd} was used. A further generalization is
suggested in order to take into account  space dependence of the
parameters describing  the effectiveness of the government
restrictions (quarantine rules) and the virus transmission
mechanism.

The LSC problem for the nonlinear equation (\ref{2-3}) was solved.
As a result, eleven inequivalent cases (up to equivalence
transformations) were identified depending on the death rate
$\mu(\tau)$ and the parameter $\gamma$, which is related to
restrictions introduced by authorities during the epidemic spread.
  As follows from Table~\ref{tab1*}, the extensions of the principal algebra
  (\ref{3-3}) are highly nontrivial and cannot be intuitively predicted
   (excepting Cases 1--3). We want to stress  that all the Lie algebras
    obtained are infinite-dimensional because the Lie symmetry operators
    involve arbitrary functions of the variable $t-\tau$.  It is rather unusual
     situation because Lie algebras of invariance of nonlinear evolution equations
      arising in real world applications typically are finite-dimensional. There
       are not many exceptions and probably the best known one is the nonlinear fast diffusion equation
\[ u_{t}=(u^{-1}u_{x})_{x}+(u^{-1}u_{y})_{y},  \]
admitting  an infinite-dimensional Lie algebra, which was firstly
identified in   \cite{na-70}. Several other examples concern
nonlinear systems of PDEs (not single equations\,!), including the
classical Navier--Stokes system. Very recently, it was proved that
two-component nonlinear evolution systems  related to Ricci flows
admit an infinite-dimensional Lie algebra \cite{lo-dimas-bo-2023,
ch-ki-24}.

In order  to construct exact solutions, the simplest form,
$\mu(\tau)={\rm const}$,  and the form arising in Case~5 of
Table~\ref{tab1*} were used. Notably, the latter form is plausible
because one takes into account the typical behaviour of the function
$\mu(\tau)$ (the death rate increases with age).  The TW type
solutions were  constructed in the case  $\mu(\tau)={\rm const}$. In
particular, arbitrary functions arising in the exact solution
(\ref{4-ad3}) were specified in order to obtain the total number of
infected individuals with  meaningful interpretation for modelling
of an epidemic (see Figs.~\ref{fig-1} and~\ref{fig-2}). In Case~5,
another type of exact solutions was constructed. It was demonstrated
that
 some solutions with the correctly-specified parameters produce
qualitatively correct distributions of infected population depending
on time and age (see Figs.~\ref{fig-3} and~\ref{fig-4}).

In a forthcoming paper, we are going to construct exact solutions in
the case of two space variables, which  is the most important from
applicability point of view.

\textbf{Acknowledgement}
R.Ch. acknowledges that this research was funded by the British
Academy's Researchers at Risk Fellowships Programme. V.D.
acknowledges that this research was supported by the National
Research Foundation of Ukraine, project 2021.01/0311.

\appendix

\section{Details for the proof of Theorem 1 }
 To simplify calculations, we consider the case of two space
variables $(x_1,x_2)=(x,y)$. Let us show that  the Lie algebra
generated by (\ref{2-9}) with the coefficients (\ref{2-16}),
generates  the infinite-parameter  Li group (\ref{a3}).

Formulae  (\ref{2-16}) with  two space variables take the form
\be\label{A-1}\ba{l} \xi^t=2\beta t+F^1\left(t-\tau\right), \
\xi^{\tau}=2\beta\tau+\beta_0,
\medskip \\
\xi^{x}=\beta x+H^{12}(t-\tau)y+G^1\left(t-\tau\right), \ \xi^{y}=\beta y-H^{12}(t-\tau)x+G^2\left(t-\tau\right),\medskip \\
\eta=-\frac{2\beta}{\gamma}u, \ \zeta=2\beta(1-\mu).\ea\ee The Lie
symmetry corresponding to the parameter $\beta$ has the form
\[X_1=2t\p_t+2\tau\p_{\tau}+x\p_x+y\p_y-\frac{2}{\gamma}u\p_u+2(1-\mu)\p_{\mu}.\]
So, according to the well-known algorithm, one needs to solve the
initial problem \be\label{A-2}\ba{l}\medskip
\frac{dt^*}{d\varepsilon_1}=2t^*, \
\frac{d\tau^*}{d\varepsilon_1}=2\tau^*, \
\frac{dx^*}{d\varepsilon_1}=x^*, \ \frac{dy^*}{d\varepsilon_1}=y^*,
\ \frac{du^*}{d\varepsilon_1}=-\frac{2}{\gamma}u^*, \ \frac{d\mu^*}{d\varepsilon_1}= 2(1-\mu^*), \\
 t^*(0)=t, \ \tau^*(0)=\tau, \ x^*(0)=x, \  y^*(0)=y, \ u^*(0)=u, \ \mu^*(0)=\mu. \
\ea\ee  Hereafter  $\varepsilon$ with  a lower subscript  means  a
group parameter. All the equations in (\ref{A-2}) are simple linear
ODEs, therefore one easily obtains: \be\label{A-3}
t^*=te^{2\varepsilon_1}, \ \tau^*=\tau e^{2\varepsilon_1}, \
x^*=xe^{\varepsilon_1}, \ y^*=ye^{\varepsilon_1}, \
 u^*=ue^{-\frac{2}{\gamma}\varepsilon_1}, \
 \mu^*=(\mu-1)e^{-2\varepsilon_1}+1,\ee
 where the group parameter $\varepsilon_1$ can be replaced by
 $\alpha=e^{\varepsilon_1}>0$.

The Lie symmetry corresponding the parameter $\beta_0$ has the form
$X_2=\p_\tau $. So, the corresponding Lie group is \be\label{A-4}
t^*=t, \ \tau^*=\tau+\varepsilon_2, \ x^*=x, \ y^*=y, \ u^*=u, \
 \mu^*=\mu.\ee

All other  Lie symmetries generated by (\ref{2-9}) with the
coefficients (\ref{A-1}) involve arbitrary functions. Consider the
simplest symmetry
\[X_F=F^1\left(t-\tau\right)\p_t,\]
where $F^1\not=0$ is an arbitrary smooth function. For each function
$F^1$, we need to solve the initial problem
\[\frac{dt^*}{d\varepsilon_F}=F^1(t^*-\tau^*), \ t^*(0)=t.\]
Obviously   $\tau^*=\tau,$ because  $
\frac{d\tau^*}{d\varepsilon_F}=0$ and $ \tau^*(0)=\tau$, therefore
\[ H(t^*-\tau)=\varepsilon_F+C,  \] i.e.
\[t^*=\tau+H^{-1}(\varepsilon_F+C),\]
where $H$  is a primary function for $\frac{1}{F^1}$ and  $H^{-1}$
is the inverse function to $H$. Finally, using the initial
condition, one obtains $ C=H(t-\tau)$, hence
\[t^*=\tau+H^{-1}(\varepsilon_F+H(t-\tau)).\]
Because there is an infinite number of  functions $F^1$, the above
expression can be rewritten as  $t^*=t+T(t-\tau)$, where
$H^{-1}(\varepsilon_F+H(t-\tau))=T(t-\tau)+t-\tau.$ Simultaneously,
the restriction $T(t-\tau)\not=-(t-\tau)$ springs up.  Thus, the
following group of ET involving an arbitrary function $T(t-\tau)$ is
derived: \be\label{A-5} t^*=t+T(t-\tau), \ \tau^*=\tau, \ x^*=x, \
y^*=y, \ u^*=u, \
 \mu^*=\mu.\ee

In a quite similar way, the equivalence transformations
corresponding to the Lie symmetries
\[X_{G1}=G^1\left(t-\tau\right)\p_x, \ X_{G2}=G^2\left(t-\tau\right)\p_y\]
and
\[X_H=H^{12}(t-\tau)y\p_x-H^{12}(t-\tau)x\p_y, \]
were constructed. As a result, the following ETs were derived
\be\label{A-6}t^*=t, \ \tau^*=\tau, \
x^*=x+\varepsilon_{G1}G^1\left(t-\tau\right), \ y^*=y, \
 u^*=u, \
 \mu^*=\mu;\ee
\be\label{A-7} t^*=t, \ \tau^*=\tau, \ x^*=x, \
y^*=y+\varepsilon_{G2}G^2\left(t-\tau\right), \
 u^*=u, \
 \mu^*=\mu;\ee
 and
\be\label{A-8}\ba{l}\medskip  t^*=t, \ \tau^*=\tau, \
x^*=x\cos\left(\varepsilon_H H^{12}(t-\tau)\right)+
y\sin\left(\varepsilon_H H^{12}(t-\tau)\right), \\
y^*=-x\sin\left(\varepsilon_H H^{12}(t-\tau)\right)+
y\cos\left(\varepsilon_H H^{12}(t-\tau)\right), \
 u^*=u, \
 \mu^*=\mu. \ea\ee
 Finally, taking a superposition of of ETs (\ref{A-3})--(\ref{A-8})
 and introducing new notations for arbitrary functions, we arrive
 at the infinite-parameter  Li group (\ref{a3}) in the case of  two space
variables.


\begin{thebibliography}{99}


\bibitem{abl-zep} Ablowitz, M.,  Zeppetella, A.: Explicit solutions of Fisher's
equation for a special wave speed. Bull. Math. Biol. \textbf{41},
835--840 (1979)

\bibitem {arrigo-15} Arrigo, D.J.: Symmetry Analysis
of Differential Equations. John Wiley \& Sons, Inc, Hoboken, N.J.
(2015)

\bibitem{a-h-b-94} Arrigo, D.J., Hill, J.M., Broadbridge, P.:
Nonclassical symmetries reductions of the linear diffusion equation
with a nonlinear source. IMA Jour. Appl. Math. \textbf{52}, 1--24
(1994)

\bibitem{bentout-21} Bentout, S., Chen, Y.,  Djilali, S.: Global Dynamics of an SEIR Model with Two Age Structures and a Nonlinear Incidence. Acta Appl. Math. \textbf{171}, 7 (2021).


\bibitem{ch-dav-2020} Cherniha, R., Davydovych, V.: A mathematical model for the COVID-19
  outbreak and its applications. Symmetry \textbf{12}, 12 pp. (2020)

  \bibitem{ch-dav-2022-cd} Cherniha, R.M.,   Davydovych, V.V.: A reaction-diffusion
system with cross-diffusion: Lie symmetry, exact solutions and their
applications in the pandemic modelling. Euro. J. Appl. Math.
\textbf{33}, 785--802 (2022)

 \bibitem{ch-du-da-2024}  Cherniha, R.,  Dutka, V.,  Davydovych, V.:
 A space distributed model and its application for modeling the COVID-19 pandemic in
 Ukraine. Symmetry \textbf{16}, 17 pp. (2024)

 \bibitem {ch-ki-24} Cherniha, R., King, J.: Nonlinear systems of PDEs admitting
infinite-dimensional Lie algebras and their connection with Ricci
flows. Stud. Appl. Math. \textbf{153}, e12737 (2024)

\bibitem {ch-se-pl-book} Cherniha, R.,  Serov, M.,   Pliukhin, O.:
Nonlinear Reaction-Diffusion-Convection Equations: Lie and
Conditional Symmetry, Exact Solutions and Their Applications.
Chapman and Hall/CRC Press, Boca Raton, FL (2018)


 \bibitem{cla-mans-94}  Clarkson, P.A., Mansfield, E.L.:
   Symmetry reductions and exact solutions of a class of
nonlinear heat equations. Phys. D \textbf{70}, \emph{70}, 250--288
(1994)


 \bibitem{da-da-ch-23} Davydovych, V., Dutka, V.,
 Cherniha, R.: Reaction-diffusion equations in mathematical models arising in epidemiology. Symmetry \textbf{15},
 2025 (2023)

\bibitem{kubo-07}  Ducrot, A.: Traveling wave solutions for a scalar age-structured
equation. Discrete Contin. Dyn. Syst. B \textbf{7}, 251--273 (2007)


\bibitem{ga-wint-92}  Gazeau, J.P.,  Winternitz, P.:  Symmetries of variable coefficient
Korteweg-de Vries equations. J. Math. Phys. \textbf{33}, (12)
4087--4102 (1992)

\bibitem{gild-ker-04} Gilding, B.H.,  Kersner, R.: Travelling Waves in Nonlinear
Reaction-Convection-Diffusion. Birkhauser Verlag, Basel (2004)

\bibitem{gopalsamy-77} Gopalsamy, K.: Age dependent population dispersion in linear
habitats. Ecol. Mod. \textbf{3}(2), 119--132 (1977)

\bibitem{gurtin-81} Gurtin, M.E., MacCamy, R.C.: Diffusion models for age-structured
populations. Math. Biosci. \textbf{54},  49 (1981)

\bibitem{gurtin-82} Gurtin, M.E., MacCamy, R.C.: Product solutions and asymptotic
behavior for age-dependent, dispersing populations. Math. Biosci.
\textbf{62},  1577 (1982)


\bibitem {ibr-tor-1991}  Ibragimov, N.H.,  Torrisi, M.,  Valenti, A.: Preliminary group classification of equations
$v_{tt}= f (x, v_x) v_{xx}+ g (x, v_x)$.
 J. Math. Phys. \textbf{32},
 2988--2995 (1991)

\bibitem{jin-09} Jin, Y., Zhao, X.Q.: Spatial dynamics of a nonlocal
periodic reaction-diffusion model with stage structure. SIAM J.
Math. Anal. \textbf{40}(6), 2496--2516 (2009)


\bibitem {kawa-tana-83} Kawahara, T., Tanaka, M.:
Interactions of traveling fronts: an exact solution of a nonlinear
diffusion equation. Phys. Lett. A \textbf{97},  311--314 (1983)

\bibitem{kingston-91}  Kingston, J.G.: On point transformations of evolution equations. J.
Phys. A \textbf{24}(14), L769--L774 (1991)

\bibitem{ki-sopho-93} Kingston, J.G.,
Sophocleous, C.:  On form-preserving point transformations of
partial differential equations. J. Phys. A \textbf{31}(6),
1597--1619 (1998)

\bibitem{kubo-91}  Kubo, M.,  Langlais, M.: Periodic solutions for
  a population dynamics problem with age-dependence and spatial structure. J. Math. Biol. \textbf{29},
  363--378 (1991)


\bibitem{lahno-2002}  Lahno, V.I.,  Spichak, S.V.,  Stohnii, V.I.: Symmetry Analysis of
 Evolutionary Equations [in Ukrainian]. Institute of Mathematics of NAS of Ukraine,
Kyiv (2002)

\bibitem{liu-21} Liu, Z., Wu, Y., Zhang, X.: Existence of periodic wave trains
for an age-structured model with diffusion. Discrete Contin. Dyn.
Syst. B \textbf{26}(12) 6117--6130 (2021)

\bibitem{lo-dimas-bo-2023} Lopez, E., Dimas, S., Bozhkov, Y.: Symmetries of Ricci flows.
Adv. Nonlinear Anal.  \textbf{12},  20230106 (2023)

\bibitem{McCluskey-16} McCluskey, C.C.:
Global stability for an SEI model of infectious disease with age
structure and immigration of infecteds.   Math. Biosci. Engineering
\textbf{13}, 381--400 (2016)

\bibitem{metz-86}
  Metz, J.A.J.,  Diekmann, O.: The Dynamics
 of Physiologically Structured Populations. Springer, New York
 (1986)


\bibitem {murray-1989}  Murray, J.D.: Mathematical Biology.  Springer, Berlin
(1989)

\bibitem{na-70} Nariboli,  G.:
  Self-similar solutions of some nonlinear equations. Appl. Scientific Res.  \textbf{22},
  449--461 (1970)

\bibitem{trucco-65} Trucco, E.: Mathematical models for cellular systems the von
Foerster equation. Part I. Bull. Math. Biophys. \textbf{27},
285--304 (1965)

\bibitem{trucco-1965} Trucco, E.: Mathematical models for cellular
systems. The von Foerster equation. Part II. Bull. Math. Biophys.
\textbf{27},  449--471 (1965)

\bibitem{van-pop-sopho-09}Vaneeva, O.O., Popovych, R.O., Sophocleous, C.: Enhanced group analysis
and exact solutions of variable coefficient semilinear
 diffusion equations with a power source. Acta Appl. Math. \textbf{106}, 1--46 (2009)

\bibitem{foerster-59} Von Foerster, H.: Some remarks on changing populations. In:
The Kinetics of Cellular Proliferation, 382--407 (1959)

\end{thebibliography}
\end{document}